\documentclass[aps,preprint,groupedaddress,amsmath,amssymb]{revtex4-2}

\usepackage{graphicx}
\usepackage{dcolumn}
\usepackage{newtxtext}
\usepackage{newtxmath}
\usepackage{natbib}
\usepackage{hyperref}
\usepackage{bm}
\hypersetup{
    colorlinks = true,
    urlcolor   = blue,
    citecolor  = black,
}

\newcommand{\RomanNumeralCaps}[1]
\linenumbers
\usepackage{amsmath}

\usepackage{ulem}

\newcommand{\clmax}{C_{L,\textrm{max}}}
\newcommand{\clmean}{C_{L,\textrm{mean}}}

\newcommand{\cdmean}{C_{D,\textrm{mean}}}
\newcommand{\dlmax}{\Delta_{\textrm{max}}}
\newcommand{\dlmean}{\Delta_{\textrm{mean}}}
\newcommand{\Rey}{Re}

\begin{document}


\title{Dynamic lift enhancement mechanism of dragonfly wing model by vortex-corrugation interaction}

\author{Yusuke Fujita}
\affiliation{Graduate school of Integrated Sciences for Life, Hiroshima University, 1-7-1, Kagamiyama, Higashi-Hiroshima, Hiroshima, 739-8521, Japan}

\author{Makoto Iima}
\email[]{iima@hiroshima-u.ac.jp}
\affiliation{Graduate school of Integrated Sciences for Life, Hiroshima University, 1-7-1, Kagamiyama, Higashi-Hiroshima, Hiroshima, 739-8521, Japan}

\date{\today}

\begin{abstract}
The wing structure of several insects, including dragonflies, is not smooth, but corrugated; its vertical cross-section consists of a connected series of line segments.
Some previous studies have reported that corrugated wings exhibit better aerodynamic performance than flat wings at low Reynolds numbers ($Re \simeq \textit{O}(10^3)$).
However, the mechanism remains unclear because of the complex wing structure and flow characteristics.
Although a complex corrugated structure modifies the aerodynamic characteristics and flow properties during unsteady wing motion, for example, leading-edge vortex (LEV) dynamics, which are key to lift enhancement in many insects; the details have not yet been studied.
In this study, we analysed the flow around a two-dimensional corrugated wing model that started impulsively by direct numerical simulations.
We focused on the period between the initial generation of LEVs and subsequent interactions before detachment.
For the flat wing, it is known that a secondary vortex with a sign opposite to that of the LEV, the lambda vortex, develops and erupts to discourage lift enhancement.
For corrugated wings, such an eruption of the lambda vortex can be suppressed by the corrugation structure, which enhances the lift.
The detailed mechanism and its dependence on the angle of attack are also discussed.
\end{abstract}

\maketitle

\section{Introduction}
\label{sec:Introduction}
Flying animals and vehicles vary significantly in their size and flow properties.
The cord-based Reynolds number ($\Rey$) for a straight flight ranges from $\textit{O}\left(10^{0}\right)$ (e.g., thrips) to $\textit{O}\left(10^{6}\right)$ (e.g., passenger planes) \citep{eldredge2019leading}.
As a result, the wing shapes also vary significantly.
Typical wings with a high $\Rey$ flyers, such as passenger planes, have smooth surfaces because the surface roughness reduces the lift coefficients \citep{abbott2012theory}. 
However, the wing surfaces of many insects such as dragonflies, cicadas, and bees are not smooth, and their Reynolds number is in the range of $\textit{O}\left(10^{2}\right)-\textit{O}\left(10^{4}\right)$ (low-$\Rey$ regime).
Their wings are composed of nerves and membranes, and their cross-sectional shapes consist of vertices (nerves) and line segments (membranes). 
The geometry of the shape can be regarded as a connection of objects with a V shape or other shapes. 
This type of wing is known as a corrugated wing \citep{dudley2002biomechanics}.

Aerodynamic studies on corrugated wings have contributed to its application in small flying robots, drones, and windmills, which are useful in low-$\Rey$ regime \citep{obata2014aerodynamic, wang2020sub, au2020effect, chahl2021biomimetic, holbert2018vertical}. 
Because insects have low muscular strength, corrugated structures are expected to possess aerodynamic advantages \citep{tanaka2010fabrication}.

We focused on the aerodynamic advantages of corrugated wings in low-$\Rey$ regime, wherein dynamic lift generation owing to flapping and the generation of leading-edge vortices are important, particularly for many insects, to achieve flight against their weight \citep{ellington84_aerod_hover_insec_flighIV, dickinson1993unsteady, ellington96_leadin_edge_vortic_insec_fligh, sane03_aerod_insec_fligh, chin2016flapping, eldredge2019leading, shyy2016aerodynamics}. 
Other unsteady lift enhancement mechanisms for insects have been reported, such as delayed (or absence of) stall, wake capture, and the clap-and-fling mechanism \citep{ellington96_leadin_edge_vortic_insec_fligh, dickinson99_wing_rotat_aerod_basis_insec_fligh, weis73_quick_estimates_flight_fitness, sane03_aerod_insec_fligh, chin2016flapping}, wherein unsteady vortex generation due to flapping and its interaction with the wing play central roles. 
Most previous studies on unsteady lift enhancement mechanisms assumed a smooth wing. 
Our fundamental motivation was the effect of corrugation on such unsteady lift enhancement mechanisms. 
Previous studies on corrugated wings have not fully investigated this problem.

Many previous aerodynamic studies on corrugated wings focused on fixed wings in uniform flow at angles of attack (AoAs) smaller than $20^\circ$, which may be related to the gliding flight \citep{buckholz86_funct_role_wing_corrug_livin_system, vargas2008computational, hu2008bioinspired, levy09_simpl_dragon_airfoil_aerod_at, levy10_param_study_simpl_dragon_airfoil, Anwer13_aerodynamic_performance_dragonfly, meng13_aerod_effec_wing_corrug_at, zhang15_aerodynamic_perform_dragon, ansari19_optim_morph_perfor_model_dragon, bomphrey16_fligh_dragon_damsel}.
The two mechanisms are described in the following two paragraphs, with the main assumption of steady flow.

The lift enhancement mechanism owing to the corrugated structure was discussed by Newman \textit{et al.} \cite{newman77_model_tests_wing_section} for $\Rey=\textit{O}(10^4)$, which is relatively high in this regime.
They used a model wing with two parts: a corrugated part on the leading edge side and a smoothed curved part on the trailing edge side. 
They suggested that the V-shapes in the corrugated wings acted as turbulators, provoking an early transition to turbulent flow to reduce the size of the separation bubble. 
This type of flow produces high lift \citep{wakeling93_dragon_aerod}. 
Levy \textit{et al.} \cite{levy09_simpl_dragon_airfoil_aerod_at} used the same model as Ref. \cite{newman77_model_tests_wing_section} and performed two-dimensional numerical simulation at smaller $\Rey$s, ($2000 \le \Rey \le 8000$) to determine the vortex separations from the corrugations to reattach and reported a drag reduction and increased flight performance.

Vargas \textit{et al.} \cite{vargas2008computational} performed two-dimensional numerical simulation with one of the corrugated models taken from a dragonfly (``profile 2'' in Ref. \cite{kesel2000_aerodyn_character_dragonfly}), for $\Rey < 10\ 000$ and concluded that the aerodynamic performance of the corrugated wing model was equivalent to or better than that of the wing with the envelope of the corrugated wing (profiled wing). 
They also reported a trap of vortices in the valleys of the V-shapes to make the overall flow resemble that around the profiled wings. 
Similar vortex traps have been observed in other experiments \citep{rees75_aerod_proper_insec_wing_section, obata2009flow} and in two- and three-dimensional numerical simulations \citep{ansari19_optim_morph_perfor_model_dragon, bomphrey16_fligh_dragon_damsel}.

The following studies were performed using a wide range of AoAs.
Rees \cite{rees75_aerod_proper_insec_wing_section} used a hoverfly-based corrugated wing model, and measured its lift coefficient $C_L$ and the drag coefficient $C_D$ for $\Rey=450, 800$ and $900$.
They reported that a corrugated wing yielded a larger $C_L$ for $\Rey=800$ than the profiled wing. 
They also reported that the flow around the wing behaved as if the wing had an envelope profile. 
Kesel \cite{kesel2000_aerodyn_character_dragonfly} used corrugated wing models based on the cross-sections of a dragonfly wing for $\Rey=7800$ and $10\ 000$.
The results show that $C_L$ for the corrugated wing model is larger than that of the flat plate for some AoAs.

Herein, we note that wings in unsteady motion can generate higher lifts at greater AoAs.
Dickinson \textit{et al.} \cite{dickinson1993unsteady} analyzed impulsively started flat wings and reported higher lifts over time (the maximum lift was recorded at AoA $=45^\circ$). 
The leading-edge vortex (LEV) dynamics are particularly important \citep{ellington96_leadin_edge_vortic_insec_fligh}.
An LEV is generated when a flat wing starts its translational motion from rest \citep{eldredge2019leading}.
The LEV increases in size and circulation by feeding the vortex sheet separated from the leading edge \citep{baik2012unsteady, rival2014characteristic}.
During this process, a stagnation point on the upper surface of the wing, generated because of the flow toward the wing surface by the LEV, slides towards the trailing edge as the LEV moves toward the trailing edge with growth.
Then, the direction of the flow on the wing surface between the stagnation point and the leading edge is reversed toward the leading edge, creating a secondary vortex of the opposite sign to the LEV, between the leading edge and the LEV.
The shape of this secondary vortex resembles the Greek character ‘$\lambda$’ and is referred to as a lambda vortex in this paper \citep{eldredge2019leading, fujita2023aerodynamic}. 
The LEV detaches from the wing after the stagnation point reaches its trailing edge. 
The lambda vortex also grows and is responsible for detaching the LEV from the wing by the eruption \citep{rival2014characteristic, widmann2015parameters}.
The lambda vortex also grows and is responsible for detaching the LEV from the wing during eruptions \citep{rival2014characteristic, widmann2015parameters}. 
After the LEV is released, the next LEV is generated.

In the case of corrugated wings, some studies have focused on dynamic performance.
Luo and Sun \cite{luo2005effects} performed three-dimensional numerical simulations of revolving corrugated wing models.
They concluded that the corrugated and flat wings exhibited almost the same performance.
Bomphrey \textit{et al.} \cite{bomphrey16_fligh_dragon_damsel} performed three-dimensional numerical simulations of a flapping wing.
They reported that the pressure in the V-shaped region of the corrugated wing was lower and concluded that this low-pressure region was related to vortex formation in the V-shaped region and to the wide distribution of the negative pressure area on the corrugated wing.
However, vortex dynamics due to corrugation have not been fully examined.

The theoretical case is also different from the steady-state aerodynamic theory, which requires uniform flow and circulation of the wing alone (Kutta-Joukowski theorem) \citep{Landau1987Fluid} and the unsteady-flapping flight mechanisms used by insects, such as delayed stall (LEV-wing interaction), rotation lift (circulation of wing owing to rotation), and wake capture (interaction among vortices generated from the edges).
Unsteady flight mechanisms have been extensively studied \citep{chin2016flapping, childress1981mechanics}; however, corrugation alternates or modifies such mechanisms.

Real-life dragonflies flaps their wings at high AoAs during most flapping periods \cite{ruppell1989kinematic}.
Thus, vortex dynamics among LEV and other vortices are critical from the aerodynamic performance.
However, in the case of corrugated wings, it is challenging to determine whether the generated vortex motion is due to the wing structure or wing motion.
Therefore, to discuss the dynamic aspects of the performance improvement of a corrugated wing, it is necessary to simplify the wing motion.

The purpose of this study was to understand the relationship between a corrugated wing structure and vortex motions. 
For this purpose, we considered a two-dimensional corrugated wing model. 
Furthermore, we simplified the wing motion and focused on unsteady lift generation by translating from rest. 
Translational motion is a principal component of wing motion, in addition to pitching and rotation \citep{wang2004role, eldredge2019leading}. 
This analysis expands our knowledge of the nonstationary mechanisms that insects use during flight \citep{dickinson1993unsteady}. 
A similar approach has been applied to analyze flat wings \citep{dickinson1993unsteady, ringuette2007role, kheradvar2007correlation, kim2011flexibility, jardin2021empirical}; however, this approach has not been performed intensively for corrugated wings. 
We previously investigated the aerodynamic properties of rapidly departing corrugated wings and found that lambda vortex collapse may improve the wing performance \citep{fujita2023aerodynamic}. 
In the present study, we conducted a more detailed investigation. 
The corrugated wing exhibited a larger lift than the flat wing in the large AoA regime. 
Moreover, the lift generation was closely related to the eruption of the lambda vortex. 
The suppression of eruptions is key to evaluating and classifying dynamic lift enhancement.

The remainder of this paper is organized as follows. 
In Sec. \ref{sec:Method}, the simulation method and its validation are discussed in detail. 
The results are presented and discussed in Sec. \ref{sec:Result}. 
The lift-generation process is divided into two time periods. 
The vortex interaction stage, in which vortices grow sufficiently to detach and interact with each other, is discussed with a particular focus on the effect of the lambda vortex. 
The results are summarized in Sec. \ref{sec: Concluding remarks}.

\section{\label{sec:Method}Method}
\subsection{\label{sec:Corrugated wing model}Corrugated wing model}

\begin{figure}
	\centerline{\includegraphics{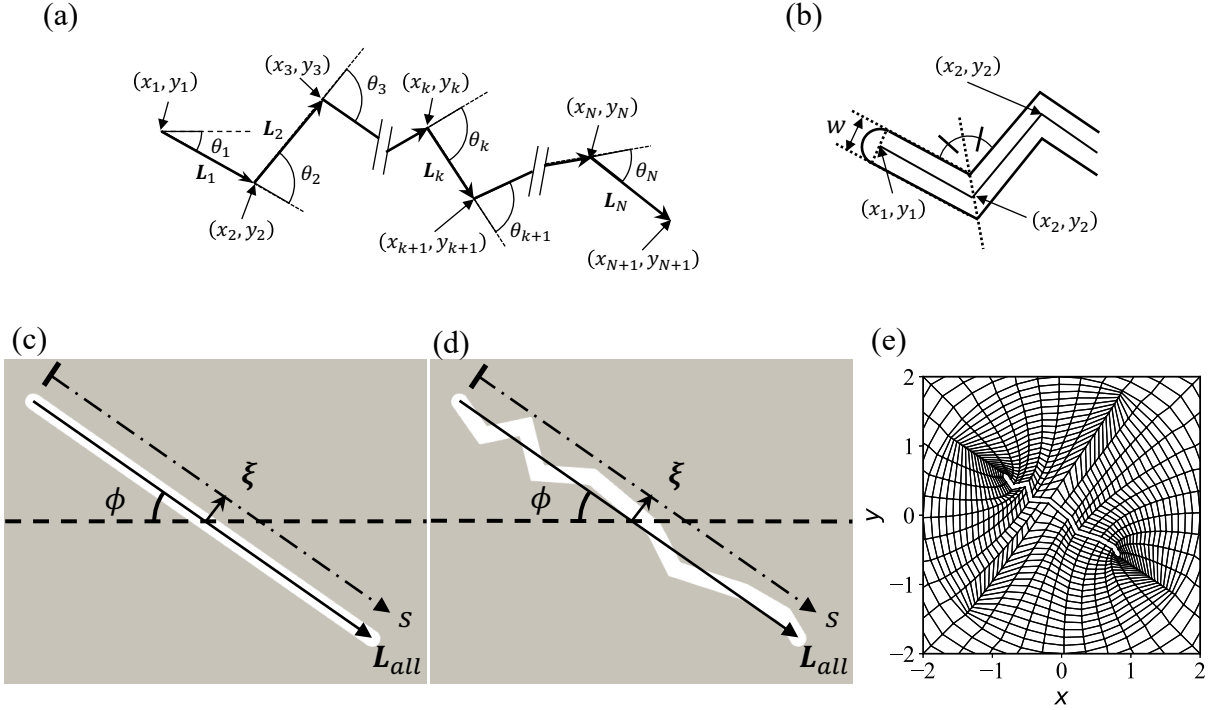}}
	\caption{\label{fig:1}Wing model descriptions. (a) Vertices $(x_k, y_k)$, adjacent displacement vectors $\bm{L}_k$, and angles $\theta_k$. (b) Line segments and thickness $w$ of the wing model. (c) Flat wing model ($\alpha = 0$). (d) Corrugated wing model ($\alpha = 1$). (e) Spectral elements around corrugated wing model ($\phi=35^\circ$).}
\end{figure}

A two-dimensional model of a corrugated wing was constructed using the dragonfly wing (\textit{Aeshna cyanea}) (Ref. \citep{kesel2000_aerodyn_character_dragonfly}, Profile 1). 
The wing model was based on a cross-section of the wing approximately 30\% from the wing base, which was approximated using $N$ line segments. 
The wing model was generated by adding thickness. 
Let us define the set of positions of the vertices and endpoints as $A=\{(x_k, y_k) \mid k=1,\cdots, N+1\}$ (Fig. \ref{fig:1} (a)).
We define the set of positions of the vertices and endpoints as $A=\{(x_k, y_k) \mid k=1,\cdots, N+1\}$ (Fig. \ref{fig:1} (a)). 
Subsequently, a set of displacement vectors is defined as $B=\{\boldsymbol{L}_k \mid \boldsymbol{L}_{k}=(x_{k+1}-x_k, y_{k+1}-y_k); k=1,\cdots, N\}$. 
The core shape of the corrugated wing model is determined using the set $C$ consisting of the line segment lengths and angles between adjacent displacement vectors, that is,
$C=\{L_k \mid L_{k}=|\boldsymbol{L}_k|; k=1,\cdots, N\} \cap \{\theta_{k+1} \mid \theta_{k+1} = \cos^{-1}\left[ \langle\boldsymbol{L}_{k}, \boldsymbol{L}_{k+1}\rangle/(L_{k} L_{k+1}) \right]; k=1,\cdots, N-1\}$,
where $\langle\boldsymbol{a},\boldsymbol{b}\rangle$ is the inner product of the vectors $\boldsymbol{a}$ and $\boldsymbol{b}$.
We defined $\theta_1$ as the angle between $x$-axis and $\bm{L}_1$ (Fig. \ref{fig:1} (a)).

After the set $C$ is determined, a family of corrugated wing shapes is characterized by introducing a shape parameter $\alpha$; the shape is obtained by replacing $\theta_k$ with $\alpha \theta_k$ in the set $C$.
In this study, we analyzed two cases: $\alpha=0$ and $1$, which corresponded to a flat plate wing (flat wing) and corrugated wing, respectively. 

The wing models were then generated by normalization, such that the wing cord length $c$ is $2$, where $c=|\bm{L}_{all}|$, and $\bm{L}_{all}=\sum_{k=1}^{N} \boldsymbol{L}_{k}$. 
The thickness $w$ of the wing was determined such that the relative thickness $w/c$ is $0.04$.
Both ends of the wing core were semicircular (Fig. \ref{fig:1} (b)). 
Figures \ref{fig:1} (c) and (d) show the wing models for $\alpha=0$ and $1$, respectively. 
AoA, $\phi$ is defined as follows $\boldsymbol{L}_{all}$ and $x$-axis, which is parallel to the uniform flow, as shown in Figs. \ref{fig:1} (c) and (d).

We also discuss the time variations in the pressure and vorticity distributions in a region near the upper surface. 
This region is characterized by a distribution along the line segment, which is a shifted vector of $\bm{L}_{all}$ by $\bm{\xi}$, where $\bm{\xi} \perp \bm{L}_{all}$ (Figs. \ref{fig:1} (c) and (d)). 
The position on the line segment is defined by $s$ ($0\leq s \leq 2$); $s=0$ and $2$ correspond to the leading and trailing edges when $\bm{\xi}=(0, 0)$, respectively. 
The amount of shift $|\bm{\xi}|$ and $s$ are nondimensionalized by the wing chord length, that is, $\xi^*=\frac{|\bm{\xi}|}{c}, s^*=\frac{s}{c}$.

\subsection{Numerical Simulation}

\begin{figure}
	\centerline{\includegraphics{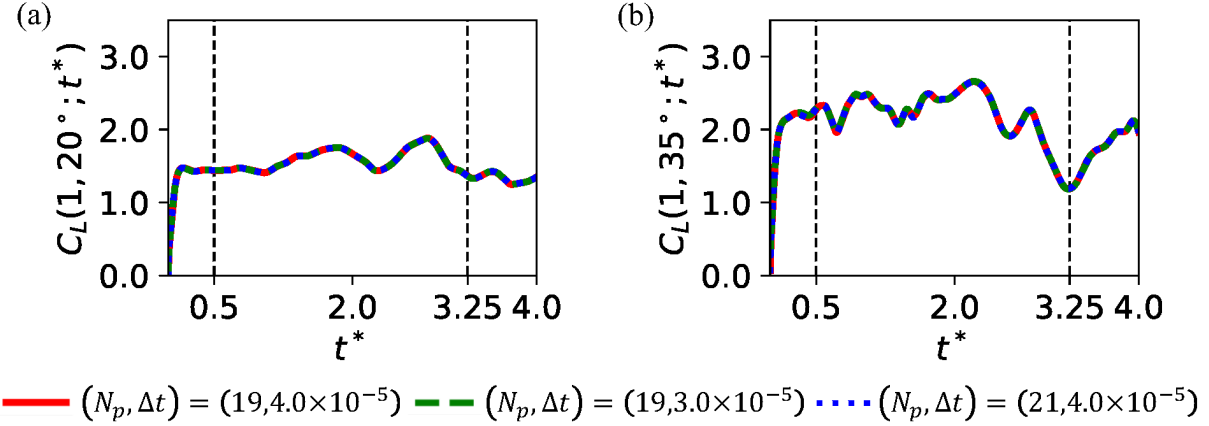}}
	\caption{\label{fig:2}Convergence of calculations for corrugated wing ($\alpha=1$). Red solid line and green broken line indicate $(N_p, \Delta t)=(19, 4.0\times 10^{-5})$ and $(19, 3.0\times 10^{-5})$, respectively. (a) $C_L(1, 20^\circ; t^*)$. (b) $C_L(1, 35^\circ; t^*)$.}
\end{figure}

 \begin{table}
 \caption{\label{tb:1}Convergence of $\clmean$ for the corrugated wing ($\alpha=1$) in the time interval $0.5<t^*<3.25$.}
 \begin{ruledtabular}
 \begin{tabular}{ccc}
    $(N_p, \Delta t)$ & $\phi=20^\circ$ & $\phi=35^\circ$ \\ \hline
    $(19, 4.0\times 10^{-5})$ & 1.5877 & 2.2151 \\
    $(19, 3.0\times 10^{-5})$ & 1.5877 & 2.2171 \\
    $(21, 4.0\times 10^{-5})$ & 1.5880 & 2.2147 \\
 \end{tabular}
 \end{ruledtabular}
 \end{table}

The dimensionless two-dimensional incompressible Navier-Stokes equations 
\begin{equation}
  \frac{\partial \bm{u}}{\partial t} + (\bm{u} \cdot \nabla) \bm{u} =
 \frac{1}{\rho}\nabla p + \nu \Delta \bm{u}, \quad
 \nabla \cdot \bm{u} = 0
 \label{eq:NS}
\end{equation}
were used to calculate the flow, where $\bm{u}=(u,v)$ is the velocity, $p$ is the pressure, $\rho$ is the air density, and $\nu$ is the kinematic viscosity. 
For the numerical calculations, we used the spectral element method, in which the computational domain was divided into a set of elements, and the physical quantities in each element were independently represented by a polynomial function with $C^0$ continuity across element boundaries \citep{karniadakis2005spectral}. 
The calculations were performed for each element using the spectral method. 
As a result, the spectral element method balanced the exponential (spectral) convergence of errors associated with global collocation methods such as pseudo-spectral methods with the geometric flexibility of traditional low-order finite element methods \citep{blackburn2019semtex}. 
The computational solver \textit{Semtex} \citep{blackburn2019semtex} was used to calculate Eq. (\ref{eq:NS}). 
The computational domain was $[-30,30] \times [-30, 30]$, and the center of the wing model was placed at $(0,0)$ (figure \ref{fig:1} (c)).

The boundary conditions of the outer sides of the computational domain, except for the right side ($x=30, -30 \le y \le 30$) was the inflow, with $\bm{u}=\boldsymbol{U}=\left(1,0\right)$.
The robust outflow conditions proposed in Ref. \cite{dong2014robust} with a smoothness parameter of $0.1$ was applied to the right side. 
In this study, we set $\nu =5 \times 10^{-4}$ such that the Reynolds number was $\Rey=4000$, based on the chord length $c$. 
At the Reynolds number $Re=1500, 4000$, the qualitative trends remain the same \citep{fujita2023aerodynamic}. 
Time step $\Delta t$ was set to $4.0\times 10^{-5}$, and the calculations were performed in the range of $0\leq t^* \leq 4$, where $t^*$ is dimensionless time $t^*=|\boldsymbol{U}|t/c$.

Figure \ref{fig:1} (e) shows the spectral elements around the corrugated wing model ($\phi=35^\circ$). 
Total number of spectral elements $N$ was $2960$, and each spectral element was discretized by $N_p \times N_p$ meshes \citep{blackburn2019semtex}. 
Therefore, $N_p$ is the number of points along the edge of each element. 
As $N_p$ increases, the total number of meshes increases.

The calculations were verified for the corrugated wing model ($\alpha=1$) at $\phi = 20^\circ$ and $35^\circ$. 
The time series of the lift coefficient $C_{L}(\alpha, \phi, t^*) =L/[\left(1/2\right)\rho c|\boldsymbol{U}|^{2}]$, where $L$ represents the lift, is shown in Figs. \ref{fig:2} (a) and (b). 
The following three computational parameter sets were investigated: $(N_p, \Delta t)=(19, 4.0\times 10^{-5})$, $(19, 3.0\times 10^{-5})$, and $(21, 4.0\times 10^{-5})$. 
All the time series agreed well with each other. 

The quantitative verification of the time averages of $C_L, \clmean$ for the corrugated wing ($\alpha=1$) are listed in Table \ref{tb:1} ($\clmean$ is defined in Eq. (\ref{eq:defs of max and mean})). 
Herein, the time interval corresponding to the intervals in Sec. \ref{sec:Result} was chosen. 
The maximum difference between the two cases $(N_p, \Delta t)=(19, 4.0\times 10^{-5})$ and $(N_p, \Delta t)=(21, 4.0\times 10^{-5})$ was $0.0004$ for $\clmean(1, 35^\circ)$ which was $0.02\%$ in the case $N_p=21$. 
The maximum difference between the two cases $(N_p, \Delta t)=(19, 4.0\times 10^{-5})$ and $(N_p, \Delta t)=(19, 3.0\times 10^{-5})$ was $0.002$ for $\clmean(1, 35^\circ)$ which was $0.09\%$ in the case $\Delta t=3.0\times 10^{-5}$.
$(N_p, \Delta t)=(19, 4.0\times 10^{-5})$ was used for all calculations in the following sections.

\section{\label{sec:Result}Result}

\begin{figure}
	\centerline{\includegraphics{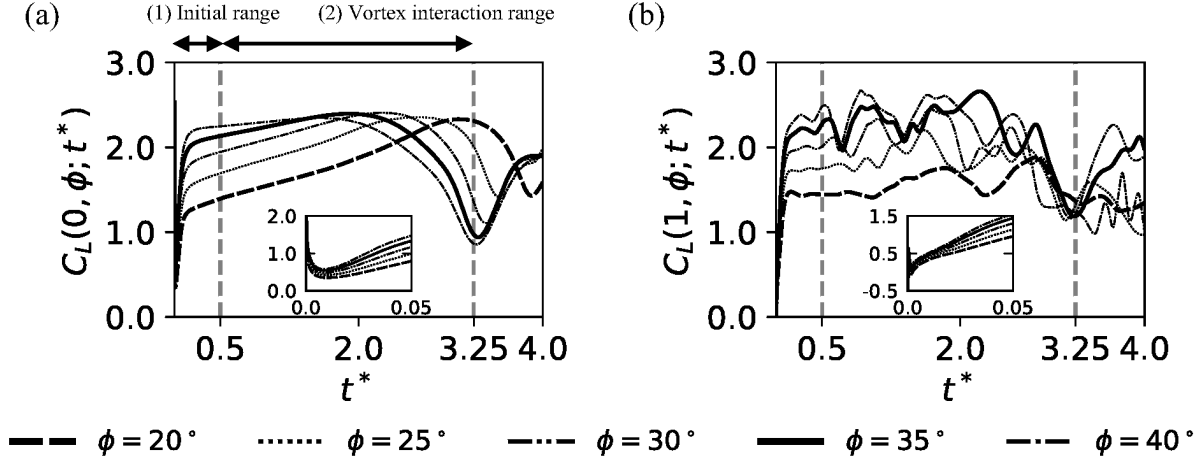}}
	\caption{Time series of the lift coefficient, $C_L(\alpha, \phi;t^*)$ ($20^\circ<\phi<40^\circ$). (a) Flat wing ($\alpha=0$). (b) Corrugated wing ($\alpha=1$).}
	\label{fig:3}
\end{figure}

\begin{figure}
	\centerline{\includegraphics{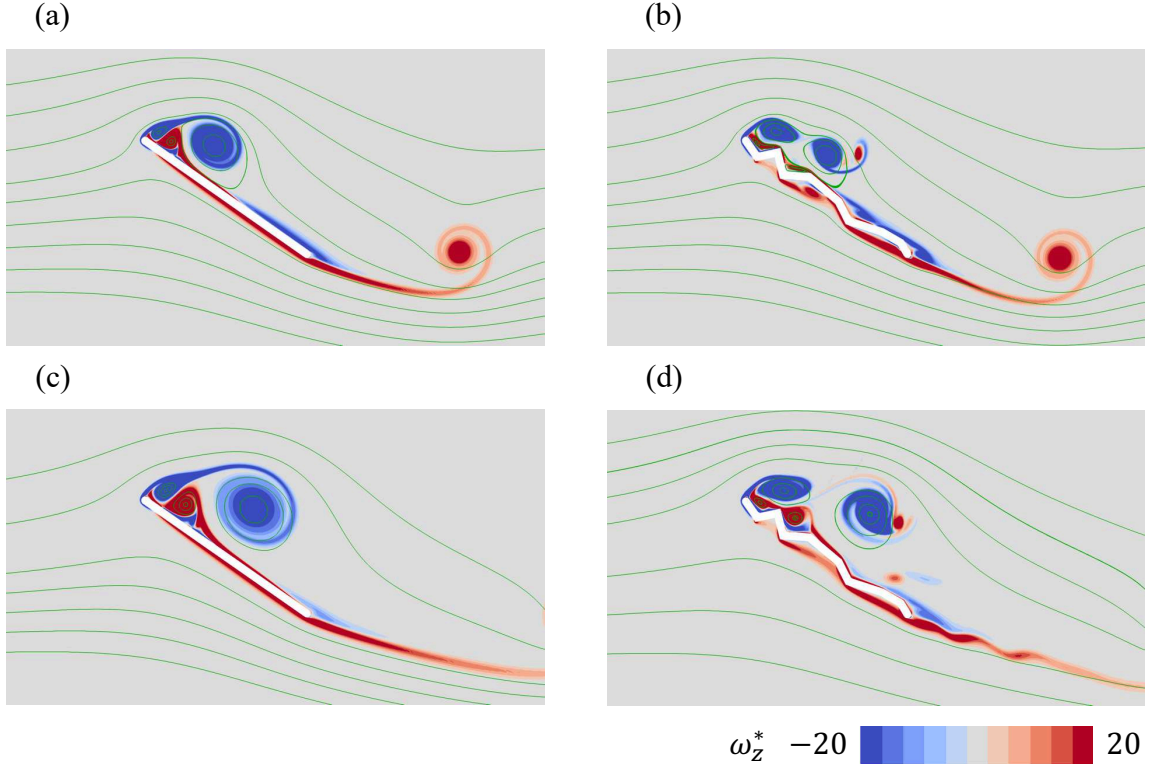}}
	\caption{\label{fig:4}Normalized vorticity fields for (a) flat wing ($\alpha=0$) for $\phi=35^\circ$ at $t^*=0.50$, 
  (b) corrugated wing ($\alpha=1$) for $\phi=35^\circ$ at $t^*=0.50$, 
  (c) flat wing ($\alpha=0$) for $\phi=35^\circ$ at $t^*=3.25$,
  and (d) corrugated wing ($\alpha=1$) for $\phi=35^\circ$ at $t^*=3.25$.}
\end{figure}

The aerodynamic wing performance was evaluated using the lift coefficient $C_L(\alpha, \phi, t^*)$ and drag coefficient $C_D(\alpha, \phi, t^*)=D/[\left(1/2\right)\rho c|\boldsymbol{U}|^{2}]$, where $D$ is the drag.
Figures \ref{fig:3} (a) and (b) show $C_L(\alpha, \phi, t^*)$ in the range $20^\circ \le \phi \le 40^\circ$ for the flat wing ($\alpha=0$) and corrugated ($\alpha=1$) wings, respectively.

In the initial range ($0< t^* \le 0.50$), $C_L(\alpha, \phi, t^*)$ shows a rapid change owing to the singularity of motion over a short period (Inset of Fig. \ref{fig:3} (a)).
Subsequently, the growth of the vortex separated from the leading and trailing edges, as well as the vertices of the corrugated structure, is observed. 
In Figs. \ref{fig:4} (a) and (b), we show the normalized vorticity fields of the flat wing and the corrugated wing, respectively ($\phi=35^\circ$ at $t^*=0.50$).
Figures \ref{fig:4} (a) and (b) show the normalized vorticity fields of the flat and corrugated wings, respectively ($\phi=35^\circ$ at $t^*=0.50$). 
Herein, the vorticity $\omega_z = \partial v/\partial x -  \partial u/\partial y$ was normalized as $\omega^{*}_{z}=\frac{\omega_z c}{|\boldsymbol{U}|}$.

In the vortex interaction range ($0.50 < t^* < 3.25$), $C_L(\alpha, \phi, t^*)$ exhibited irregular oscillations for the corrugated wing ($\alpha=1$). 
This range corresponded to the development of the LEV and secondary vortices generated on the upper wing surface, including vortex break-ups and interactions among the vortices, before leaving the vortices in the vicinity of the wing.
The end of the range was determined by a decrease in $C_L(\alpha, 35^\circ, t^*)$ (Fig. \ref{fig:3} and Figs. \ref{fig:4} (c) and (d)); however, this range included the maximum lift time for the other cases during computation.

To characterize these ranges, we defined the maximum and mean lift coefficients in the time interval $[a,b]$, $C_{L,\textrm{max}}(\alpha, \phi)$ and $C_{L,\textrm{mean}}(\alpha, \phi)$, respectively, as follows:
\begin{equation}
\begin{split}
  C_{L,\textrm{max}}(\alpha, \phi)&=\underset{a \le t \le b}{\textrm{max}}
     C_L(\alpha, \phi;t),\; \\
  C_{L,\textrm{mean}}(\alpha, \phi)&=\frac{1}{b-a}
  \int_{a}^{b} C_L(\alpha, \phi;t) dt,
  \label{eq:defs of max and mean}
\end{split}
\end{equation}
and similar definitions for $\cdmean$.

These values were used to characterize the aerodynamic wing performance, and we discussed the relationship between these values and the flow field, that is, the pressure field, vorticity field, and flow speed field. 
The pressure was evaluated using the pressure coefficient, $C_{p}=\frac{p-p_{\infty}}{\left(1/2\right)\rho c|\boldsymbol{U}|^{2}}$, where $p$ is the pressure and $p_{\infty}$ is the inflow pressure (at positions $x=-30$). 
The flow speed was normalized by a uniform flow, $u^*=\frac{|\boldsymbol{u}|}{|\boldsymbol{U}|}$, where $|\boldsymbol{u}|$ is the flow speed.

The major developments and interactions of the vortices were observed in the vortex interaction range ($0.50 < t^* < 3.25$). 
Therefore, in the following $(a, b)=(0.5, 3.25)$ in Eqs. (\ref{eq:defs of max and mean}).

\subsection{\label{sec: Wing performance}Wing performance}

\begin{figure}
	\centerline{\includegraphics{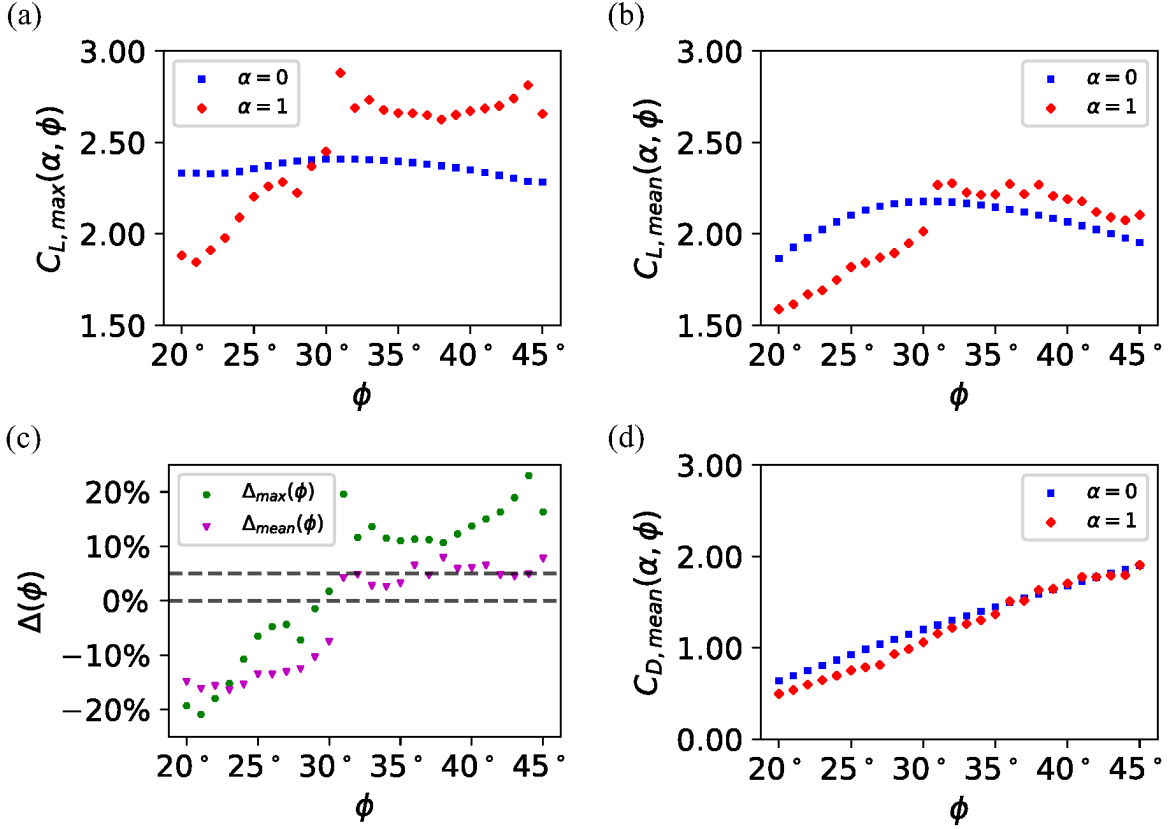}}
	\caption{\label{fig:5}(a) $\clmax$ during $0.5< t^* < 3.25$ for the flat wing (blue; $\alpha=0$) and corrugated wing (red; $\alpha=1$).
	(b) $\clmean$ during $0.5< t^* < 3.25$ for the flat wing (blue; $\alpha=0$) and corrugated wing (red; $\alpha=1$).
	(c) $\dlmax$ and $\dlmean$ during $0.5< t^* < 3.25$. Vertical axis is in a percentage form.
	(d) $\cdmean$ during $0.5< t^* < 3.25$ for the flat wing (blue; $\alpha=0$) and corrugated wing (red; $\alpha=1$).}
\end{figure}

Figures \ref{fig:5} (a) and (b) show that the values of $\clmax$ and $\clmean$ for the corrugated wing are smaller than those of the flat wing when $\phi<30^\circ$.
However, the values of $\clmax$ and $\clmean$ for the corrugated wing are larger than those of the flat wing when $\phi>30^\circ$.
These graphs suggest that the lift coefficients of the corrugated and flat wings interchange at a critical value $\phi \simeq 30^\circ$. 
In particular, the performance of the corrugated wing clearly improves when $30^\circ < \phi$ ($\le 45^\circ$); the mean value of $\Delta_{\textrm{max}}$ over this range is $0.14$ and the mean value of $\Delta_{\textrm{mean}}$ over this range is $0.05$. 
Because the mean value of $\Delta_\textrm{mean}$ over the deteriorated case ($20^{\circ}\leq$) $\phi < 30^{\circ}$ was $-0.14$, the improvement between these cases is $22.1\%$.

In Fig. \ref{fig:5} (d), $\cdmean$ is plotted against $\phi$.
No clear transitions are observed. 
When $\phi > 35^\circ$, the values of $\cdmean$ for the corrugated and flat wings are almost the same, whereas $\cdmean$ for the corrugated wing is smaller than that for the flat plate wing at $\phi < 35^\circ$.
Therefore, the qualitative tendency of the lift-to-drag ratio $L/D$ is similar to that of $C_L$. 
Therefore, we discuss $C_L$ as wing performance hereafter.

\begin{figure}
	\centerline{\includegraphics{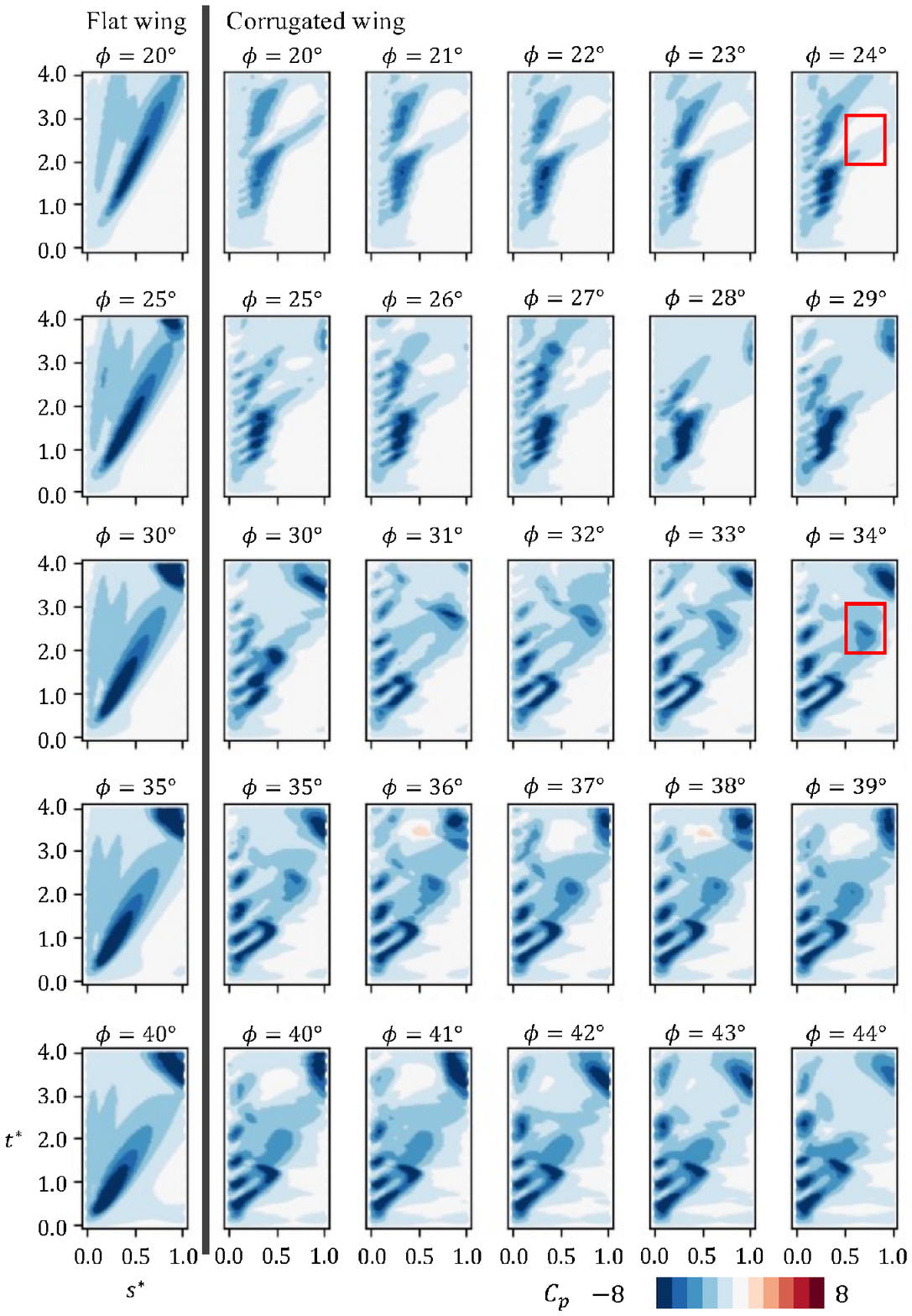}}
	\caption{\label{fig:6}Spatiotemporal distributions of $C_p$. $\xi^*=0.15$. $s^*$-axis and $t^*$-axis are the horizontal and vertical axes, respectively.}
\end{figure}

The sign of $\dlmax$ and $\dlmean$ can be related to the pressure fields above the wing, as discussed below. 
Figure \ref{fig:6} shows the spatiotemporal distributions of the pressure on $(t^*, s^*), (0\leq t^*\leq 4, 0 \leq s^* \leq 1)$ for $\xi^*=0.15$. 
The leftmost column shows the distributions of the flat-wing cases in $\phi=20^\circ, 25^\circ, 30^\circ, 35^\circ$ and $40^\circ$. 
The five columns on the right show the distributions for the corrugated wing in $20^\circ \leq \phi \leq 44^\circ$.

When wing performance is improved ($30^\circ \leq \phi$; cf. Fig. \ref{fig:6}), low-pressure regions appear near the leading-edge side periodically with a period of about $0.6$. 
Such periodic regions are absent in corrugated wings without improved wing performance and in flat wings. 
Additionally, low-pressure regions appear periodically on the trailing edge side ($0.5 \leq s^*$) during $2<t^*<3.25$.
This trend is also observed for $\xi^*=0.2$ (data not shown); it is insensitive to $\xi^*$.

\subsection{\label{sec:Lift enhancement case: time series of the lift coefficient and instantaneous pressure and flow field}Lift enhancement case: instantaneous pressure and flow field}

\begin{figure}
	\centerline{\includegraphics{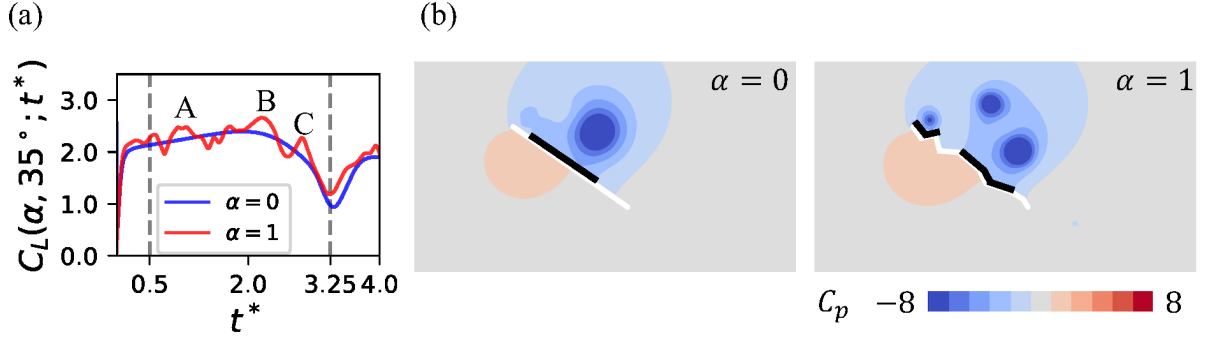}}
	\caption{\label{fig:7}(a) Time-series of lift coefficient (blue: flat wing ($\alpha =0$), red: corrugated wing ($\alpha =1$); $\phi=35^\circ$). (b) Snapshots of the pressure field around the wings at $t^*=t^*_\textrm{max}$.}
\end{figure}

We selected the case $\phi=35^\circ$ as the typical case for the improvement of corrugated wing performance because the values of $\clmax$, and $\clmean$ were similar for $33^\circ \leq \phi \leq 43^\circ$ (Figs. \ref{fig:5} (a) and (b)).
Therefore, the vortex dynamics described below can be considered a typical mechanism of dynamic lift enhancement.

Figure \ref{fig:7} (a) shows the lift coefficients $C_L(\alpha, 35^\circ; t^*)$ for the flat wing ($\alpha=0$), and corrugated wing ($\alpha=1$). 
A single maximum is recorded at $t^* \simeq 1.89$ for the flat wing, whereas multiple maxima are recorded for the corrugated wing owing to an oscillation at approximately $C_L(0,35^\circ;t^*)$. 
The oscillations of $C_L(1,35^\circ;t^*)$ suggest a complex interaction between the vortices and the wing. 
For the following discussion, three major maxima, A, B, and C, are designated for $C_L(1,35^\circ; t^*)$.
$t^*_{\textrm{max}}$ is defined as the time that yields the greatest value of $C_L(\alpha, 35^\circ; t^*)$, $t^*_{\textrm{max}}= 1.89$ for the flat wing ($\alpha=0$) and $t^*_{\textrm{max}}= 2.21$ for the corrugated wing ($\alpha=1$ corresponding to maximum B).

Figure \ref{fig:7} (b) shows the pressure fields at $t^*=t^*_{\textrm{max}}$. 
On the upper surface of the wing, the low-pressure regions of the corrugated wing are distributed over a wider range than those of the flat wing (regions where $C_p \leq -\frac{24}{11}$ are indicated by thick black lines). 
On the lower surface of the wing, the corrugated wing has a wider high-pressure region than the flat wing, which is similar to the initial range. 
Similar pressure distributions were observed at maxima A and C (data not shown).

\subsection{\label{sec: Vortex dynamics of lift generation: role of the lambda vortex}Vortex dynamics of lift generation: role of the lambda vortex}

\begin{figure}
	\centerline{\includegraphics{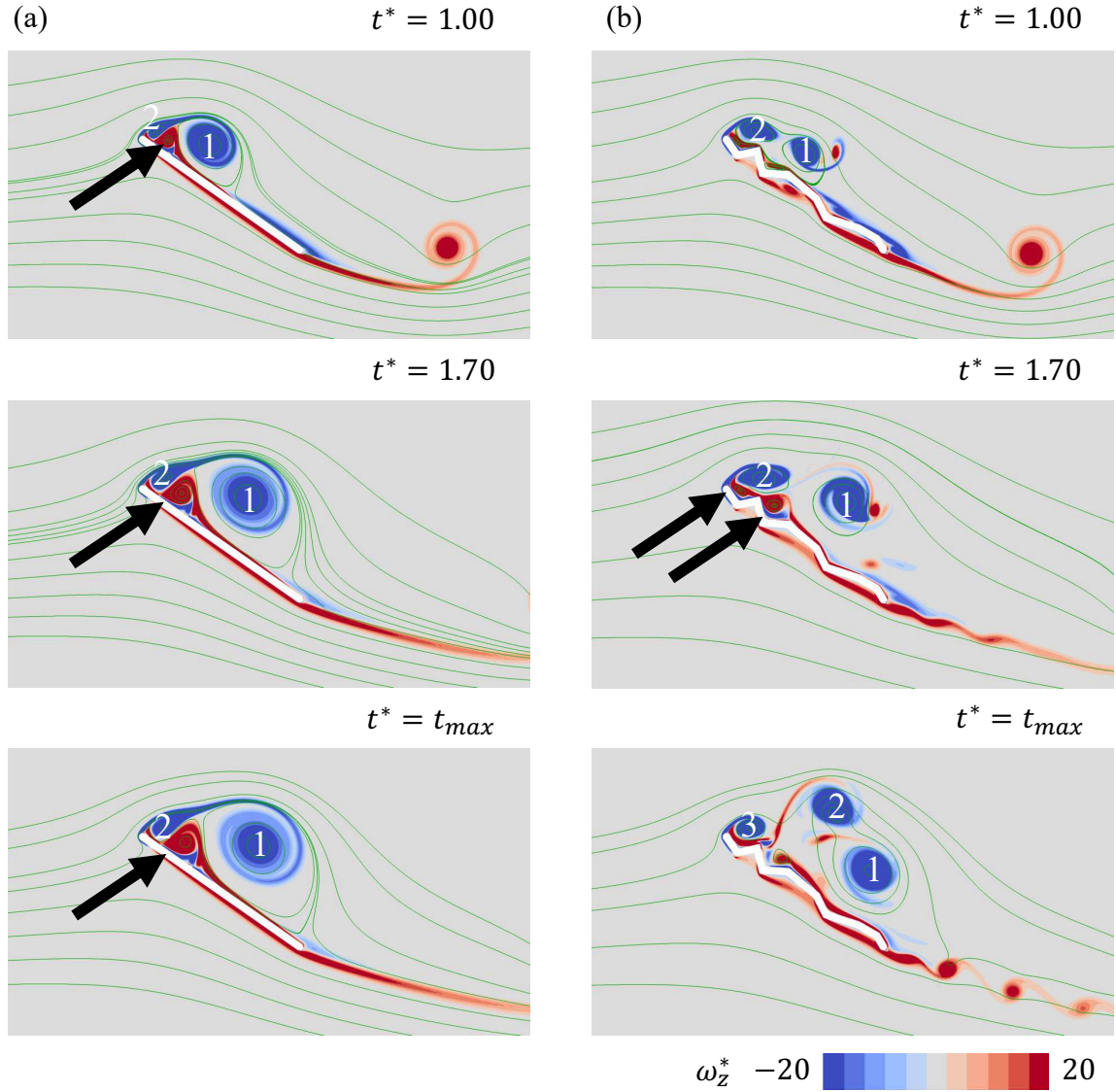}}
	\caption{\label{fig:8}Snapshots of the normalized vorticity fields ($\phi=35^\circ$) for (a) flat plate wing; $t^*=1.00$, $t^*=1.70$ and $t^*=t_{\textrm{max}}$, and (b) corrugated wing; $t^*=1.00$, $t^*=1.70$ and $t^*=t_{\textrm{max}}$.}
\end{figure}

The key vortex dynamics for the improving wing performance is as follows.

Figure \ref{fig:8} (a) shows the vorticity fields around a flat wing ($\alpha=0$). 
Two snapshots at $t=1.00$ and $1.70$ are selected to explain the vortex dynamics before $t=t_{\textrm{max}}$.
At $t=1.00$, two vortices with negative signs were formed from the leading edge (labeled 1 and 2), and a secondary vortex with a positive sign was sandwiched between them, as indicated by the arrow. 
The secondary vortex is called a lambda vortex \citep{rival2014characteristic, eldredge2019leading, dickinson1993unsteady}.

Vortex 1 continues to grow by feeding a vorticity-containing mass \citep{rival2014characteristic} as the value of $C_L$ continues to increase until $C_L$ records the maximum. 
During this period, these three vortices develop without changing their relative configurations, moving the center of vortex 1 away from the wing. 
Concurrently, the stagnation point owing to the flow pushing on the surface of the wing slides downstream to approximately reach the trailing edge. 
This result is consistent with those of previous studies (Ref. \cite{rival2014characteristic, eldredge2019leading}). 
In terms of pressure, the low-pressure region at the center of vortex 1 moved away from the wing. 
Consequently, the lift generation process is monotonic, and the (local) maximum occurs only once in the vortex interaction range.

However, the behavior of the lambda vortex on the corrugated wing is significantly different from that of the flat wing because of the irregular surface structure; that is, the lambda vortex collapses, splits into several smaller vortices owing to the corrugated structures, and gets stuck in the V-shaped regions of the wing (Fig. \ref{fig:8} (b), $t^*=1.70$, indicated by arrows). 
Consequently, the lambda vortex behavior replaces the unsteady vortex dynamics that characterize insect flight mechanisms, as described below.

Figure \ref{fig:8} (b) shows the vorticity field around the corrugated wing ($\alpha=1$). 
At $t^*= 1.00$, two vortices with negative signs are formed near the leading edge (labeled 1 and 2). 
These vortices were originally separated from the leading edge as in the case of a flat wing ($\alpha=0$). 
Here, vortex 1 is detached from the vortex sheet connected to the leading edge, as is vortex 2 soon after $t^*=1.70$. 
Therefore, vortices 1 and 2 act as independent vortices, which is a major difference from the case of the flat wing. 
Consequently, the relative positions of the vortices at $t^*=t_\textrm{max}$ is significantly different from that in the case of the flat wing (Fig. \ref{fig:8} (b)).

Consequently, vortex 2, which is nearest to vortex 1, advects vortex 1 downward to the wing surface. 
Unlike in the case of the flat wing, the lambda vortex does not contribute to the motion of vortex 1 owing to the collapse. 
The proximity of vortex 1 results in a wider high-pressure region at $t^*=t^*_{\textrm{max}}$ (maximum B) (Fig. \ref{fig:8} (b); c.f. Fig. \ref{fig:7} (b) for the pressure field).

\begin{figure}
	\centerline{\includegraphics{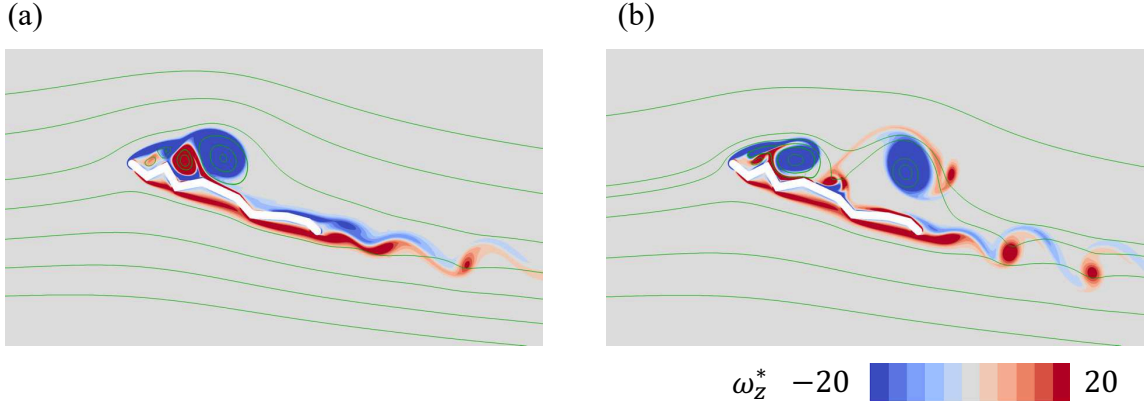}}
	\caption{\label{fig:9}Typical snapshots of the normalized vorticity fields around the corrugated wing ($\alpha=1, \phi=20^\circ$). (a) $t^*=2.15$. (b) $t^*= 2.81=t^*_{\textrm{max}}$.}
\end{figure}

When wing performance is not improved ($\phi\le 30^\circ$), the lambda vortex grows on the upper surface of the wing (Fig. \ref{fig:9} (a)). 
The lambda vortex then stretches to interfere with the larger LEV, preventing it from being pulled to the wing (Fig. \ref{fig:9} (b)).

\begin{figure}
	\centerline{\includegraphics{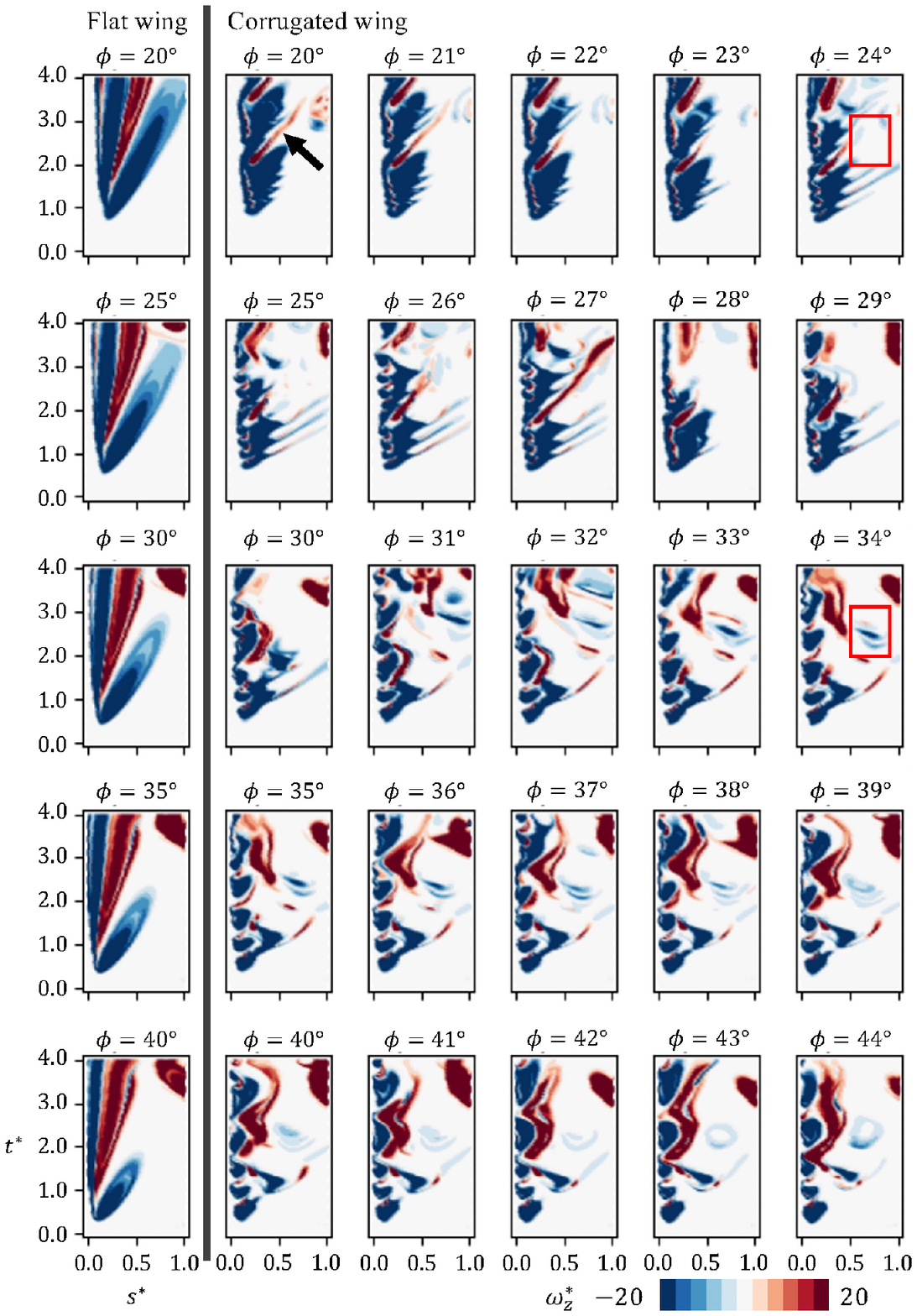}}
	\caption{\label{fig:10}Spatio-temporal distributions of the vorticity field. $\xi^*=0.15$. $s^*$-axis and $t^*$-axis are the horizontal and vertical axes, respectively.}
\end{figure}

In both cases, with and without an improvement in the performance of the corrugated wing, the vortex dynamics explained above were common. 
Figure \ref{fig:10} shows the spatiotemporal distribution of vorticity on $(t^*, s^*), (0\leq t^* \leq 4, 0\leq s^* \leq 1)$ for $\xi^*=0.15$. 
The leftmost column shows the distributions for the flat wing cases at $\phi=20^\circ, 25^\circ, 30^\circ, 35^\circ$ and $40^\circ$. 
The remaining five columns show the same for the corrugated wing at $20^\circ \leq \phi \leq 44^\circ$.

\begin{figure}
	\centerline{\includegraphics{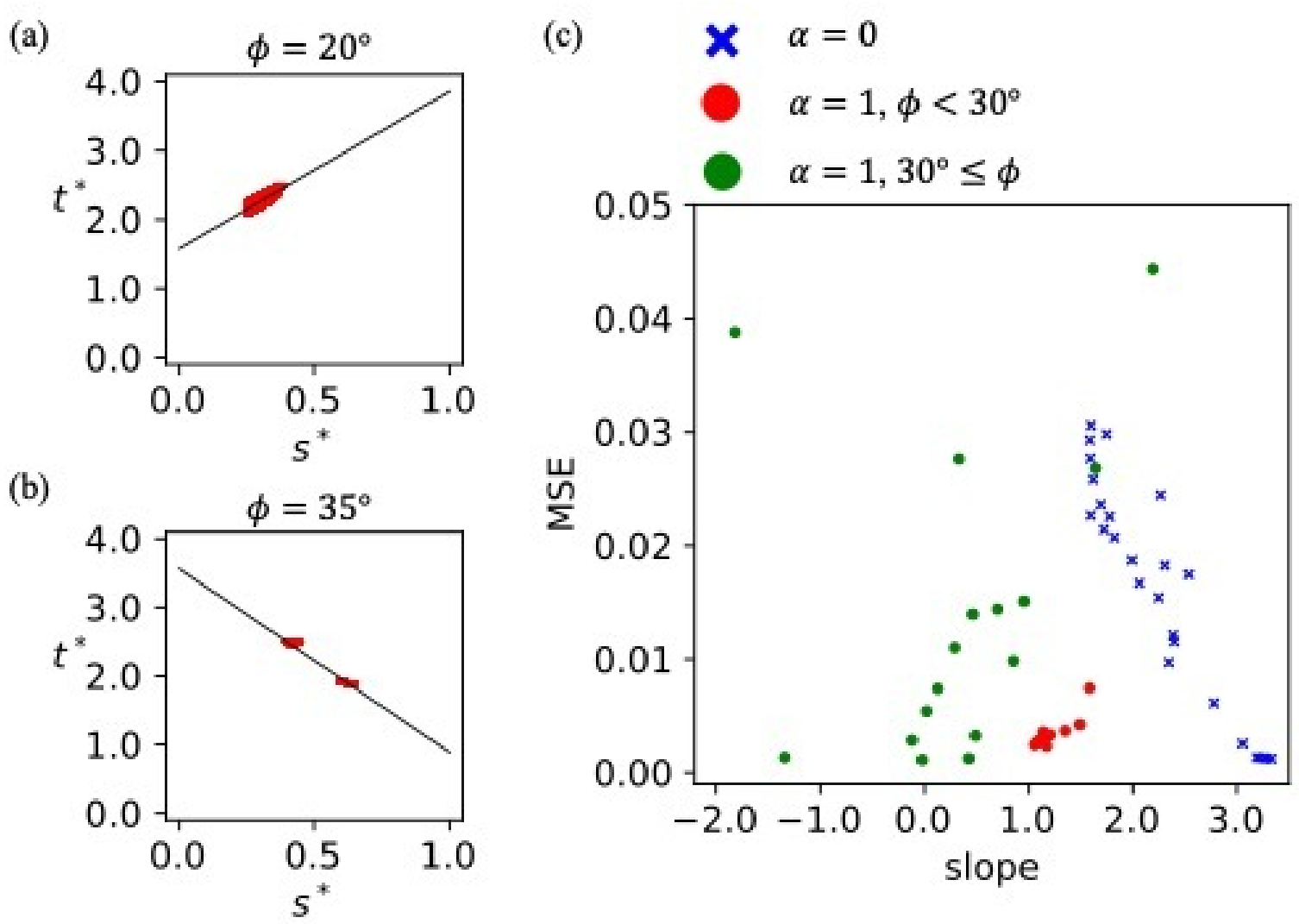}}
	\caption{\label{fig:11}Extraction of lambda vortex eruptions. (a) $\phi=20^\circ$. (b) $\phi=35^\circ$. (c) Results of linear regression for the distribution of the lambda vortex. Phase diagram of slope and mean squares error.}
\end{figure}

The eruption of the lambda vortex corresponds to a positively signed vortex recorded near $t^*=2, s^*=0.25$; c.f. arrow in $\phi=20^\circ$ in Fig. \ref{fig:10}.
Subsequently, the eruption causes the vortex to move toward the trailing edge with time. 
To quantify the eruption, a vorticity field satisfying $s^*>0.25, 1.5<t^*<2.5, \omega^*_z\geq20$ was extracted to perform linear fitting using the least-squares method. 
For example, the extracted regions and fitted lines for the $\phi=20^\circ$ and $35^\circ$ are shown in Figs. \ref{fig:11} (a) and (b), respectively. 
The mean squared error (MSE) of the fitting and slope of the regression line are plotted in Fig. \ref{fig:11} (c).

In the case of no improvement in the wing performance (Fig. \ref{fig:11} (c); red circle), the slope is concentrated around $1$ and the MSE is small and localized. 
This implies that the vortex with a positive sign moves to the trailing edge, corresponding to a lambda vortex eruption. 
When the wing performance is improved (Fig. \ref{fig:11} (c); green circles), the slope is small or the variation is large, and the result depends on $\phi$.
No clear lambda-vortex eruption is identified. 

For the flat wing (Fig. \ref{fig:11} (c); blue cross), the slope is concentrated in a range larger than $1$. 
This also indicates that a vortex with a positive sign has moved to the trailing edge in that region, corresponding to a lambda-vortex eruption.

For corrugated wings with $30^\circ\leq\phi$, vortices with negative signs are periodically generated at the leading-edge side, as shown in Fig. \ref{fig:10}. 
This corresponds to the pressure-field results shown in Fig. \ref{fig:6}. 
This trend is also observed for $\xi^*=0.1$; it is insensitive to $\xi^*$.

\subsection{\label{sec:Time-average}Mean behaviour}

\begin{figure}
	\centerline{\includegraphics{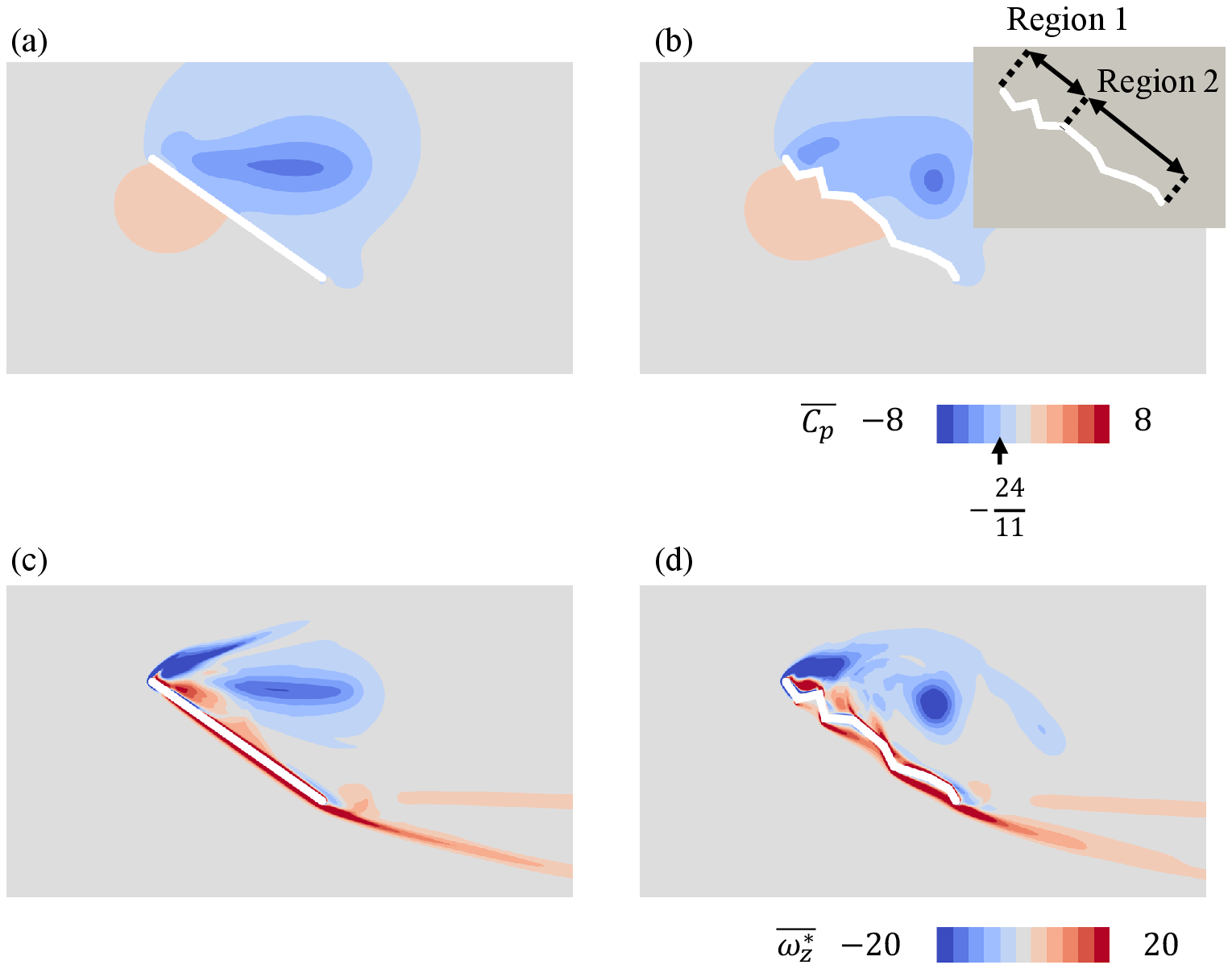}}
	\caption{\label{fig:12} Mean pressure fields around (a) flat wing ($\alpha=0$, $\phi=35^\circ$) and (b) corrugated wing ($\alpha=1$, $\phi=35^\circ$), and the definition of the region around the corrugated wing model for understanding the flow characteristics. Mean normalized vorticity fields around (c) flat wing ($\alpha=0$, $\phi=35^\circ$) and (d) corrugated wing ($\alpha=1$, $\phi=35^\circ$).}
\end{figure}

\begin{figure}
	\centerline{\includegraphics{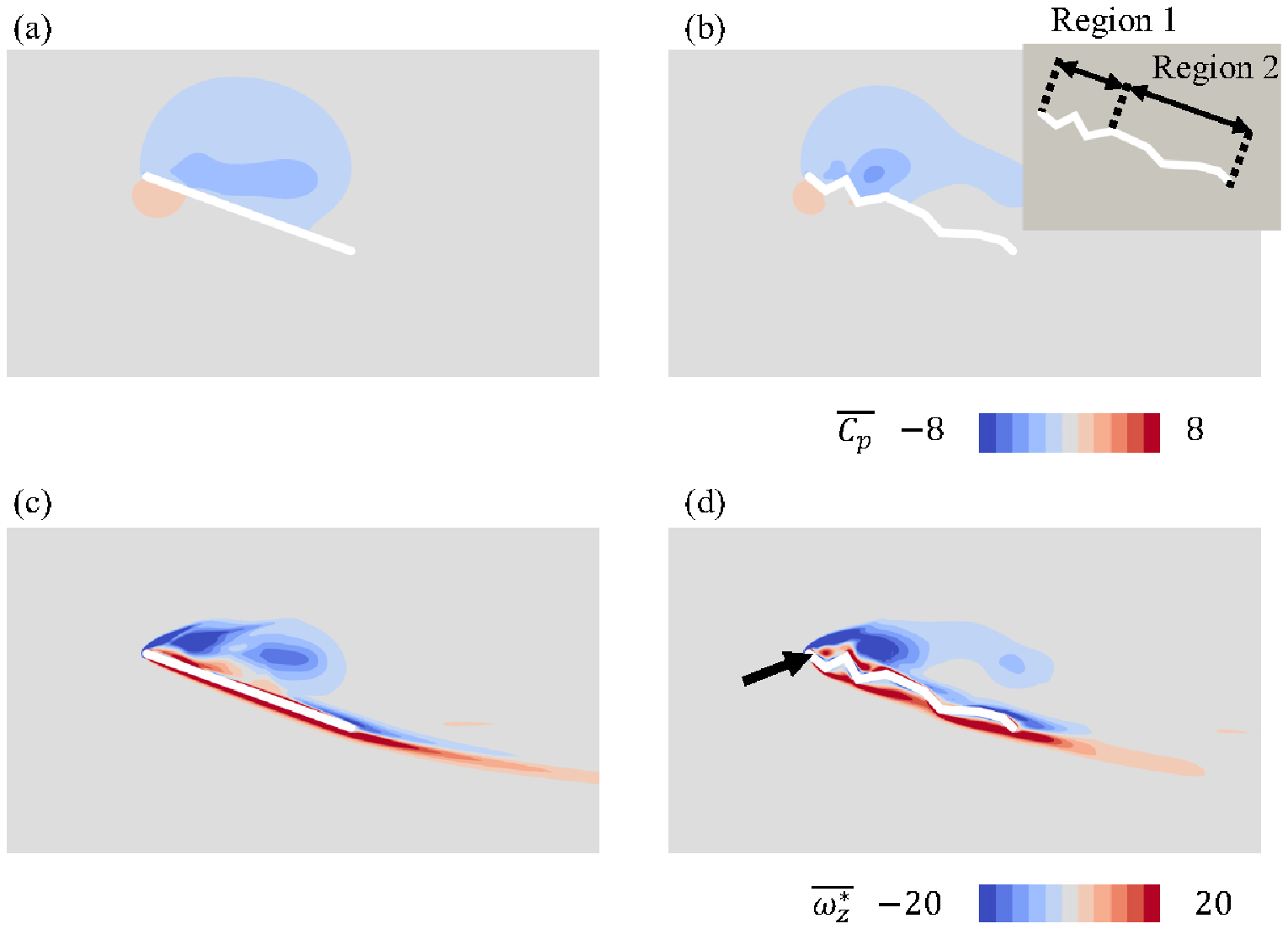}}
	\caption{\label{fig:13}Mean pressure fields around (a) flat wing ($\alpha=0$, $\phi=20^\circ$) and (b) corrugated wing ($\alpha=1$, $\phi=20^\circ$), and the definition of the region around the corrugated wing model for understanding the flow characteristics. Mean normalized vorticity fields around (c) flat wing ($\alpha=0$, $\phi=20^\circ$) and (d) corrugated wing ($\alpha=1$, $\phi=20^\circ$).}
\end{figure}

In this subsection, we focus on the mean flow behavior to discuss the lift enhancement mechanism of the corrugated wing over the time interval of the vortex interaction range. 
We compared the following two cases: $\phi=35^\circ$ for the lift enhancement, and $\phi=20^\circ$ when the lift is not enhanced. 
The mean pressure and vortex fields are shown in Fig. \ref{fig:12} (case $\phi=35^\circ$) and Fig. \ref{fig:13} (case $\phi=20^\circ$). 
For the discussion below, we define the following two regions: region on the upper side near the leading edge (Region 1) and that of the rest of the wing (Region 2; Insets of Figs. \ref{fig:12} (b) and \ref{fig:13} (b)).
We define the following two regions, the region on the upper side near the leading edge (Region 1), and that on the rest of the wing (Region 2; insets of Figs. \ref{fig:12} (b) and \ref{fig:13} (b)).
Region 1 contains two V-shaped regions and the boundary is defined by $(x_5, y_5)$.

We discuss the behaviors in Regions 1 and 2 separately. 
In Region 1, the negative pressure region on the corrugated wing was wider than that on the flat plate when $\phi=35^\circ$ (Figs. \ref{fig:12} (a) and (b); for example, the region where $C_p \leq -\frac{24}{11}$). 
This negative pressure region corresponded to a round vortex region with a positive sign stuck in the V-shaped region (Fig. \ref{fig:12} (d)), which was caused by the collapse of the lambda vortex. 
The details are presented in Sec. \ref{sec: Vortex dynamics near the leading-edge: how low pressure region is generated}. 
However, for the case $\phi=20^\circ$, the negative pressure distributions for both wings were similar in this region (Figs. \ref{fig:13} (a) and (b)).
Stronger negative-pressure region (where $C_p \leq -\frac{24}{11}$) did not cover the wing surface (Fig. \ref{fig:13} (b)) because the round vortex was smaller (Fig. \ref{fig:13} (d); arrow).

In Region 2, the pressure distributions on both wings were similar for $\phi=35^\circ$ (Fig. \ref{fig:12} (a) and (b)).
The vorticity distribution formed a round shape for the corrugated wing, and the LEV remained in this region for a certain time interval (Fig. \ref{fig:12} (d)). 
In contrast, the vorticity region for the flat plate wing was elongated as a result of LEV development in a similar manner (Fig. \ref{fig:12} (c)).

As for $\phi=20^\circ$, the negative pressure region on the flat plate wing was wider than that on the corrugated wing (Figs. \ref{fig:13} (a) and (b)), which was consistent with the fact that the value of $\dlmean$ was negative (cf. Fig. \ref{fig:5} (c)). 
For a corrugated wing, the round vorticity region observed in the case $\phi=35^\circ$ shifted to the leading-edge side, and no distinct vorticity region was observed in this region (Fig. \ref{fig:13} (d)), which explained the narrow negative pressure region (Fig. \ref{fig:13} (b)). 
In contrast, the vorticity region for the flat plate was relatively close to the wing surface, and the negative-pressure region was maintained (Fig. \ref{fig:13} (a)).

In summary, the mean lift enhancement was analyzed by focusing on these two regions. 
On the lower surface of the wing, the flow was almost steady and the `profiled wing' image was valid for all the investigated $\phi$s, although previous studies have been limited to smaller AoAs \citep{vargas2008computational, levy09_simpl_dragon_airfoil_aerod_at}.
However, on the upper surface, vortex dynamics played a decisive role in evaluating lift generation. 
When the corrugated wing generated a larger lift, the LEV remained near the wing, which required the elimination of interference by the lambda vortex. 
The corrugation broke the lambda vortex to become stuck in V-shaped regions. 
However, the detailed dynamics near the leading edge require further discussion, as discussed in the following subsection.

\subsection{\label{sec: Vortex dynamics near the leading-edge: how low pressure region is generated}Vortex dynamics near the leading-edge: how low pressure region is generated}

\begin{figure}
	\centerline{\includegraphics{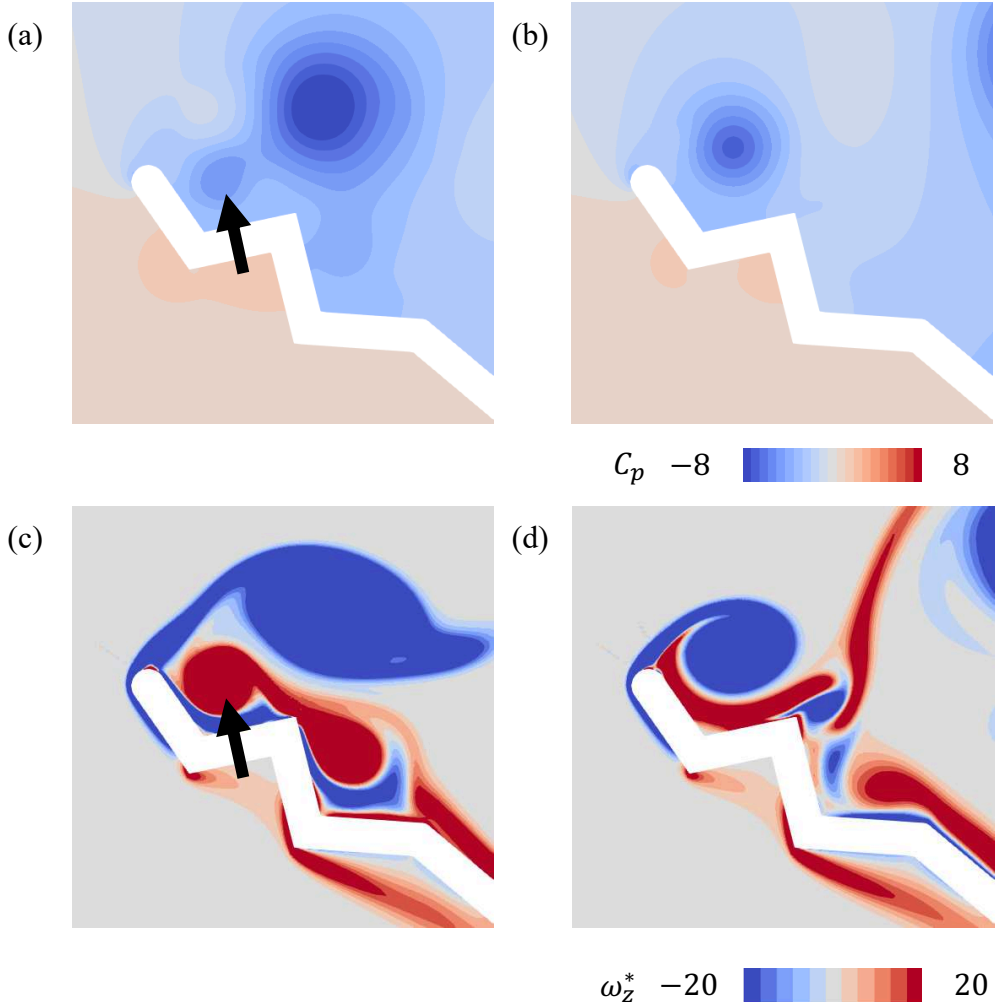}}
	\caption{\label{fig:14}Typical snapshots of the flow field near the leading edge of the corrugated wing ($\alpha=1, \phi=35^\circ$). (a) $t^*=1.80$. (b) $t^*=2.15$.}
\end{figure}

In this subsection, we explain the generation of low pressure in the V-shaped region due to the lambda vortex collapse and LEV in the corrugated wing (cf. Sec. \ref{sec:Time-average}).

On the upper surface of the wing, the corrugated wing has a strong negative pressure area in the first V-shaped region counted from the leading edge. 
This low-pressure area does not form in the second V-shaped region. 
However, for the flat wing, a strong negative-pressure area is present at a certain distance from the leading edge. 
In both cases, the negative-pressure region on the wing owing to the proximity of the vortex, (as discussed in Sec. \ref{sec:Lift enhancement case: time series of the lift coefficient and instantaneous pressure and flow field} (Fig. \ref{fig:7} (b))) did not appear in the mean field (Fig. \ref{fig:12} (b)).
In contrast, the negative-pressure region in the first V-shape remains a snapshot when the maximum lift is recorded (Figs. \ref{fig:7} (b) and \ref{fig:8}).

Figure \ref{fig:14} shows the flow fields for a corrugated wing ($\alpha=1$) at $t^*=1.80$ and $2.15$. 
These particular times are chosen to evaluate wing performance near the leading edge in terms of the mean flow field such that the two typical vortex dynamics explained below are clearly visible.

Figures \ref{fig:14} (a) and (c) show the pressure and vorticity fields at $t^*=1.80$, respectively. 
A vortex with a negative sign is formed from the leading edge, and a secondary vortex with a positive sign, generated by the collapse of the lambda vortex, is formed between the wing and the vortex. 
Accordingly, a negative-pressure region is formed in the first V-shaped region (Figs. \ref{fig:14} (a) and (c), indicated by the arrow).

Figures \ref{fig:14} (b) and (d) show the pressure, vorticity, and velocity fields, respectively, at $t^*=2.15$. 
Here, a vortex with a positive sign is squashed. 
Accordingly, the squashed region correspond to the low-speed region. 
A vortex with a positive sign in the leading-edge concavity forms a dead-water region and is transmitted, and the negative pressure created by the LEV acts on the wing surface.

As described above, the dynamics in the first V-shaped region is not stationary, but dynamic. 
Nonetheless, a low-pressure region is generated because both round vortices exhibit positive and negative signs, and they contribute to the low-pressure region. 
The periodic generation of vortices with both signs is also observed in Figs. \ref{fig:6} and \ref{fig:10}.
This process persists after time averaging, as shown in Fig. \ref{fig:12}.

\section{\label{sec: Concluding remarks}Concluding remarks}
In this study, the flow around a two-dimensional corrugated wing was analyzed using direct numerical calculations at $\Rey=4000$, and the wing performance was compared with that of a flat wing. 
The performance of the corrugated wing was better when the AoA was greater than $30^\circ$.

The uneven structure of the corrugated wing generates an unsteady lift owing to complex flow structures and vortex motions. 
Herein, we discuss several lift-enhancement mechanisms owing to the uneven structure of a single corrugated wing model of a dragonfly.

The first is the pressure reduction on the upper side of the wing owing to the interactions of the vortices detached from the LEV. 
The detachment and formation of vortices results from the collapse of the lambda vortex. 
The second is the dynamic generation of a low-pressure region in the V-shaped structure near the leading edge on the upper side of the wing. 
Here, the collapsed lambda vortices and LEV stuck in the V-shaped region near the leading edge form a negative pressure region, thereby generating an averaged negative pressure area. 
To the best of our knowledge, these mechanisms are dynamic and have not been reported elsewhere.

Such mechanisms can be used for novel wing shape designs, particularly in the low-Reynolds-number regime corresponding to insect flight. 
This is similar to the passive drag reduction of turbulent flow; the drag can be reduced by approximately 8\% by simply changing the surface shape \citep{dean2010shark}.
However, for further understanding, the relationship between wing shape and aerodynamic performance should be studied in more detail. 
These mechanisms are used to analyze various corrugation patterns and flows (for example, different $\Rey$s).
The authors also studied an inverted corrugated wing model to obtain similar results \citep{fujita2023aerodynamic}.
Additional details have been reported elsewhere.

Additionally, the dependence on the Reynolds number is important. 
We have reported that the qualitative trends remain broadly the same at the Reynolds number, $Re=1500, 4000$ \citep{fujita2023aerodynamic}.
However, as the Reynolds number decreases, the vortex motion may differ from that shown here because of the viscous effects. 
Further details are reported elsewhere.

In this study, we considered two-dimensional models. 
However, this study focused on the aerodynamics of insect flights, in which the flow is typically two-dimensional. 
If these results are expanded to a three-dimensional system, we expect to gain more practical knowledge for understanding insect flights and their application in the industry. 
Investigations in three-dimensional space will be the subject of future research. 
We can gain a clearer understanding by resolving these issues.

\begin{acknowledgments}
This work was supported by JSPS KAKENHI (19K03671, 21J22603, 21H05303) and the SECOM Science and Research Foundation.
\end{acknowledgments}


\bibliography{draft.bib}

\end{document}